**Protein Structure and Evolutionary History Determine Sequence Space Topology**

Boris E. Shakhnovich[1], Eric Deeds[2], Charles Delisi[1] and Eugene Shakhnovich[3]

*[1]Bioinformatics Program, Boston University*

*44 Cummington Street, Boston MA, 02215*

*[2]Department of Molecular and Cellular Biology, Harvard University*

*[3]Department of Chemistry and Chemical Biology, Harvard University*

*12 Oxford Street, Cambridge MA, 02138*

**Abstract**

Understanding the observed variability in the number of homologs of a gene is a very important, unsolved problem that has broad implications for research into co-evolution of structure and function, gene duplication, pseudogene formation and possibly for emerging diseases. Here we attempt to define and elucidate the reasons behind this observed unevenness in sequence space. We present evidence that sequence variability and functional diversity of a gene or fold family is influenced by certain quantitative characteristics of the protein structure that reflect potential for sequence plasticity i.e. the ability to accept mutation without losing thermodynamic stability. We identify a structural feature of a protein domain – contact density - that serves as a structural determinant of entropy in sequence space, i.e. ability of a protein to accept mutations without destroying the fold (also known as fold designability). We show that the (log) of the average gene family size exhibits statistical correlation ($R^2>0.9$.) with the contact density of its three-dimensional structure. We present evidence that the sizes of individual gene families are influenced also by their evolutionary history e.g. the amount of time the gene family was in existence. We further show that our observed statistical correlation between gene family size and designability of the structure is valid on many levels of evolutionary divergence i.e. not only for closely related gene but also for less related fold families.

**Introduction:**

Gene family and domain fold family sizes are known to vary widely[1; 2; 3; 4; 5; 6; 7] – from orphans (families that have only a single member) to considerably populated sets of far-diverged homologs. The observed variability in the number and divergence of gene family members raises many questions e.g. which genetic mechanisms and evolutionary dynamics could have led to the observed unevenness. Evolutionary biologists have proposed models designed to explain these size distributions (which often follow power laws [4; 7; 8; 9] while assuming no inherent physical differences between gene families from the outset.[4; 8; 10; 11] However, many of these models are overly abstract to adequately explain family size distributions in a constructive manner that relates specific features of gene families with their reported size. Neither do these models provide explicit insights into the mechanistic details that might explain observed differences. On the other hand, some researchers have hypothesized that the heterogeneity in family size is due to an underlying distribution of biological or physical properties [3; 12; 13; 14; 15; 16] of proteins encoded by gene sequences, but until now such properties could only be hypothetically characterized for a limited class of simplified two dimensional and three dimensional lattice models.

In particular, in a recent study Taverna and Goldstein[12] analyzed the contribution from various factors such as evolutionary history and fold designability to the development of uneven protein family sizes in simplified 2-dimensional lattice models. These authors modeled several scenarios of evolution and demonstrated that more "designable" structures indeed feature more populated (or overpopulated) sequence families. Interestingly, they find that the relationship between designability of a structure (defined in their model as a number of sequences that can have non-degenerate ground state in that structure) and the size of the family exhibits a noticeable scatter indicative of the influence of evolutionary history on the observable outcome[12].

Recent successes in structural genomics and bioinformatics provide a wealth of data for statistical analysis of the distributions of gene family sizes of real proteins with known structures. On the other hand, our research has increased our understanding of the structural determinants of protein designability[17; 18; 19] and has made it possible to analyze the structural features of real protein domains that might be responsible for the observed inequality of gene family sizes. Obtaining new insights into the relative roles of physical and biological factors that contribute to the genesis of modern gene families may bring us closer to a greater understanding of the natural history of protein domains.

From a biological perspective, we may hypothesize that gene family size is at least in part influenced by functional constraints related to the number of different but perhaps related functions needed by the cell[20]. For example, some functions such as kinase activity have varied specificities within a relatively small number of sequence mutations[21] while others such as globins have much less functional flexibility despite, in some cases, substantial sequence divergence.[22] From a physical perspective, the potential of a gene to obtain new function upon duplication may depend on its ability to accept mutations without destroying the three-dimensional structure of a protein domain that it encodes. In this work we will focus mostly on the effect of the physical constraints imposed on the structure encoded by sequences of the gene family. We will show that variability in these constraints represents difference in potential for sequence diversity of gene families. This effect can be observed for real families both on average and in the case of specific families taking into account their differential time of evolution.

**Building PDUG:**

In order to consider sequence, structure and function information in a unified, systematic way, we define both gene families and fold families quantitatively using the Protein Domain Universe Graph (PDUG)[8]. The PDUG is a graph where nodes are sets of closely related sequences folding into structurally characterized domains[23; 24] and edges are connections between the nodes that are based on structure comparison. (Fig.1). The sequences of the domain structures inside each node exhibit less than 25% identity to other representative sequences in PDUG. Thus, nodes in PDUG are a set of *representative* structures. To include all available sequence data we incorporate both SWISS-PROT[25] and NRDB90[26] databases. This enables us to calculate the size of the gene family by employing all available sequence data and at the same time discounting database bias. We use NRDB to calculate gene family size and SWISS-PROT in combination with Inter-Pro[27] to calculate functional divergence for every domain.(see Methods)  We obtain structures from the Dali Domain Dictionary[24] and use BLAST[28] and DALI[29] sequence and structure comparison tools (see Methods). Thus the size of the gene family as represented on PDUG is the number of non-redundant sequences from NRDB that are highly homologous to the representative structure of the domain inside the node.

Using this PDUG formalism, we can define a gene family based on micro-evolutionary considerations: the PDUG represents the *variability* accessible to a given gene upon mutation, whether that variability occurs in sequence, function or structure space. Unlike other definitions of gene families [23; 30], we would like our definition to be entirely local, i.e. the definition should be made with respect to a *particular gene*. The gene family of a gene is therefore all the immediate sequence neighbors of that gene that do not significantly alter the structure [31; 32]. By our construction of the PDUG, a gene family is represented by sequences within a single PDUG node. Similarly, the fold family of a structure is defined as all the structural neighbors of that domain on PDUG (Fig.1). By defining the cutoff value for sequence or structure comparison (see Methods) we can control the allowed variance of that particular attribute. Thus we implicitly control the time scale of evolutionary divergence over which we calculate structure-function determinants.

**The Role of Designability:**

Our first task is to determine what, if any, physical factors are responsible for the variability in gene family size. To this end we define an inherent *structural* characteristic related to the number of sequences that a structure can accommodate without loss of thermodynamic stability i.e. we employ a structural determinant of designability[13]. This feature has been previously hypothesized[3; 13; 14] to be one of the key influences responsible for over-representation of some folds over others. Recent analysis[18] suggested that structures with greater values of traces of powers of their contact matrices (CM) (i.e. $Tr[CM]^2$, $Tr[CM]^4$ etc) are predicted to be more designable[18](see Methods). Sequence space Monte Carlo[18] calculations for simple lattice models show that this characteristic of a structure does indeed correlate strongly with its designability which we define as logarithm of the number of sequences that are stable in the structure.

The physical explanation for the correlation between traces of powers of the CM (a structural feature) and sequence entropy (i.e. designability) follows from the fact that the trace of powers of the CM reflects topological characteristics of the network of contacts within the structure. For example, the trace of $CM^2$ simply gives the total number of contacts (or equivalently the total number of two step, self-returning walks) and the trace of $CM^4$ reflects the number of length-4 closed loops in the system and so on. One may also note that certain closed sets of contacts allow for optimal placement of amino acids that interact very favourably. For example, if four amino acids that strongly attract each other are folded into an architecture where they all interact favourably (e.g. on four corners of a square, see Fig.3) this formation represents a greater contribution to the stability of the overall structure than configurations in which the same four amino acids are arranged linearly or in cases where the last of the contacts is out of the contact range (Fig.3). Such optimal placement of several strongly interacting amino acids allows more sequences to be folded into the structure by relaxing energy constraints *for the rest of the sequence*. Thus structures that provide certain features, such as availability of long closed loops of interactions and higher density of contacts per residue, are expected to be able to accommodate a wider variety of different sequences. This qualitative argument is similar in spirit to derivation of Boltzmann distribution in Statistical Mechanics[33] and similar to the justification for the ''Boltzmann device'' used in the derivation of knowledge-based potentials[3; 34] for the study of protein folding and prediction of ligand binding energies.

For this study we employ the trace of the second order of the contact matrix normalized by chain length as a simplest approximation for designability. This quantity, known as the contact density (CD), is proportional to the number of contacts per amino acid residue (see Methods): it corresponds to the lowest second-order term in the expansion of Eq.1. A designability criterion, at this level of approximation has been considered earlier by several authors[17; 19], and these studies predicted that the number of contacts, along with other factors such as dispersion of interaction energies as well as the proportion of long and short-range contacts in a structure may play an important role in determining the designability of a structure.

We thus calculate the CD for every representative domain structure in PDUG as a measure of the designability of that node. We then define a gene family as the set of sequences with more than 25% identity to the sequence of the crystallized structure of the domain excluding close sequence homologues using NRDB90[26]. Clearly, this calculation is predicated on the assumption that Swiss-Prot and NRDB represents a fair estimate of the variability inside each gene family. Remarkably, we observe that there is a marked positive correlation between a domain's designability calculated via CD and the average gene family size (Fig.3a). However, we note that the observed correlation, while very pronounced, is nonetheless statistical in nature: each point in Fig.3 is a bin in (log) family size that contains 100-250 domains with a distribution of CD values, and the distributions in different bins overlap. Regardless of this observation we find that, on average, gene families that encode more designable protein structures are statistically the ones that perform more varied functions[27], encode more sequences and therefore constitute larger families.

We perform a similar analysis on distantly related gene families as defined through the structural comparisons within the PDUG. To this end we take the structural neighbourhood of a given domain to be all those domains that are connected to it by an edge on the PDUG[8] (Fig.1). Physically this means that all domains that are structurally but not sequentially similar to a given domain (beyond some threshold Z-score value) are included in this structural neighbourhood (see Methods). We then look at the correlation between the combined size of the "family" of gene sequences that fold into structures belonging to the same structural neighbourhood on the PDUG and the average CD for that neighbourhood. Fig. 3b shows that the average CD, which serves as a proxy for average designability of a structural neighbourhood, itself correlates with the (log) of the gene sequence family size of that neighborhood. Together, Fig.3a and Fig.3b show that gene family size and designability (as approximated by CD) correlate on average, across various scales of evolutionary distance. This could indicate that designability affects large sequence-structure spaces spanning not only sequence but also structural diversity. From an evolutionary standpoint, this may indicate that domains with higher CD diverge to produce other high-designability domain structures.

Since these observations of correlations between designability and gene family size are statistical in nature we want to comment on the robustness of the reported results. There are two issues to consider: the variability of contact density (CD) for structures within gene families and the robustness in the calculation of the mean number of sequences for all gene families in each bin. To address these concerns we first calculate the intra-family deviation in CD for each gene family on PDUG (See Methods). While the points in Fig 3a,b show mean values of the CD for the representative domains (nodes on PDUG), we also include estimates of the deviation in CD taking into account sequences inside gene families with solved structures (i.e. domains that have sequence homology to the representative domain). In order to calculate this deviation, we take all solved structures for domains with sequence homology to the representative domain and calculate the standard deviation of CD inside each gene family. We then calculate the average standard deviation in every bin of Fig 3a,b. The deviation is shown as CD-axis error bars on Fig 3a,b. It is apparent from the size of the error bars that the deviation in CD within each gene family is relatively small, on the order of .05 or less. Indeed, as expected, the intra-family dispersion deviation of CD gets smaller as average contact density increases. The CD deviation ranges from .01 at CD=4.8 to .06 at CD=3.8 in Fig 3a. The deviation is much smaller when considering domains inside structural neighbourhoods, the deviation falls to be on the order of .001. Next, we calculate the possible error in the calculation of the mean in the size of the gene family for each bin. This quantity is proportional to the square root of the number of observations in the bin, according to Central Limit Theorem. We include this as the gene family size axis error bars in Fig 3. It is worth noting that this measures the deviation of the mean over all gene families belonging to a given bin only and does not reflect the scatter of the distribution inside the bin that is considered in detail separately later. Clearly the consideration of both these errors is small enough so that it does not affect the conclusions drawn from Fig 3a,b.

We also determine how gene family size is related to the diversity of functions that family performs. We define the *functional* determinant of a gene family as entropy in function space. When we calculate this measure in the context of PDUG, we utilize Gene Ontology (GO)[35] to define the functional variability (functional flexibility score or FFS) of a set of genes (see Methods). FFS is a measure of the total amount of information needed to describe all the functionality of a gene family. Perhaps not surprisingly, FFS statistically correlates with CD (Fig 4). This is not surprising because FFS statistically correlates with the total number of sequences in a gene family (data not shown). However, this analysis serves two purposes. First the correlation of FFS and CD shows that designability directly affects the underlying biology of the domain. Domains with low CD have a much lower chance of performing many different functions. Secondly, this serves as a corroboration of the previous result using a different database, annotation method, and a completely different measure of sequence variability. Finally, the correlation of FFS instead of just simply calculations of gene family size ensures that we measure entropy on sequences that are sufficiently diverged to yield different functions thus minimizing the effect of database bias.

**The Role of Evolution:**

While statistical correlations of gene family sizes and FFS with CD are highly significant, how predictive are they when it comes to calculations of gene family size for a particular domain? To answer this question, we present a scatter plot of gene family size versus CD that shows all domains in the PDUG (Fig.5). The scatter is very significant and it is clear that CD is hardly a predictor of gene family size for an each domain. This is perhaps not surprising given that other factors may have influenced gene family sizes. A natural possibility that has also been observed in lattice simulations [12] is that the evolutionary history of protein domains may have influenced their gene family sizes. The more time a gene family has to diverge the larger the gene family because there is a higher chance of finding a suitable sequence mutation.

Understanding the evolutionary history of all the protein domains on the PDUG requires construction of the most parsimonious scenario for protein structure evolution, a complex proposition[36] that is beyond the scope of this work. The simplest construction that still yields useful information is the delineation of the very old domains from the "less very old domains". Any domain that exists in every proteome within a given set can be placed in the last universal common ancestor (LUCA) of that set representing domains that were the predecessors of all others[36]. If any such domain were not placed in the LUCA, multiple independent discovery (or horizontal transfer) events would be required to explain the occurrence of this domain in all proteomes. The "extra" evolution involved in this case would result in a less parsimonious scenario. Inclusion of other domains is more probabilistic and depends on the exact form and method of parsimony construction used.[36]

We thus define the structural content of the prokaryotic LUCA to be the intersection between all the 59 prokaryotic structural proteomes available at the time of this study, i.e. a domain is included if and only if it is present in all 59 prokaryotic genomes. The presence was determined using two way stringent BLAST with cutoff of 1e-6. This intersection that we call LUCA consists of 108 structures, representing roughly 3% of the domains in the PDUG. Of these 108 domains, 56% represent $\alpha/\beta$ folds, 19% are $\alpha+\beta$, 13% are all $\beta$ and 12% are all $\alpha$. (The list of LUCA domains is available as supplementary material) Although the structural content of the actual prokaryotic LUCA may be a superset of these domains, the 108 nonetheless represent a minimal set of structures that were most likely to have been present before the divergence of the bacteria and archaea.

We may thus highlight the LUCA domains on the scatter plot in Fig.5. Two observations are immediately apparent. First, LUCA domains clearly feature greater CD's, suggesting that ''first'' domains were more designable (difference of means .48, t-test P-value is <1e-14). Secondly, even at equal CD (designability) with their younger counterparts, LUCA domains feature greater family sizes, on average 116 more members (red points are markedly shifted towards higher family size in Fig.5, P-value < 1e-14). This observation provides evidence that, as simulations on simple lattice models suggest [9; 12], designability is only the *potential* for larger family size that has to be coupled with other mitigating factors for a full understanding of the evolutionary history of that domain. For two domains with the same CD but differing times of divergence, the domain with the longer divergence time will most likely have more sequence members. However, we can see the importance of designability even within the LUCA domains by noting that higher CD domains exhibit higher gene family size within LUCA. This is consistent with out theory that given the same amount of time for divergence, higher CD domains will have larger sequence families.

**Discussion**

In this paper, we presented evidence that across widely varying evolutionary distances there are significant statistical correlations between structural designability, functional flexibility and gene family size. The statistical nature of these observations is obvious from the scatter plot presented in Fig.5. We have found that this scatter may be explained, at least in partly, by variations in the evolutionary history [37] of protein domains. Because of this, neither CD nor any other proxy calculation of designability can be used as a *predictor* of gene family size. As shown by simulation[12], designability represents only the "potential" for sequence entropy allowed by a structure. The actual size depends not only on the potential but also on the amount of time that evolution had to explore the sequence space around that structure. This, in part, reconciles the very strong correlation of the means observed in Fig 3 and the significant scatter of the specific observations in Fig 5.

While we believe that these results are illuminating, we must mention several caveats. Using CD as a proxy for entropy in sequence space is an approximation that assumes, among other things, that protein energetics may be correctly represented in contact form and that the second-order approximation of Eq.2 is sufficient to capture the designability of a structure. An additional and perhaps more interesting caveat to consider is that the "designability principle'' in its canonical form assumes equilibrium in sequence space in which all structures take full advantage of their designability potential and that this fact is reflected in the data. Consideration of phylogeny clearly shows that this is not an entirely valid assumption. On the other extreme, several dynamic divergent evolution models predict uneven fold populations without assuming any structural preferences due to designability[8], positing that gene family sizes may be due to pure chance in the complex natural history of protein domains. Our observations are not inconsistent with divergent evolution. In fact, we have done simulations that indicate that a combination of divergent evolution models and designability yield a stunning correspondence with observed phenomena. [38]

In this work we clearly see that domains with low CD are most likely to represent smaller size families while more designable, higher CD domains may exhibit both large and small family sizes. This is exactly what one would expect from the interplay of historical and physical factors: while physical constraints impose upper bounds on sizes of families of low-CD domains, more designable domains may exhibit greater family sizes if they are "old'' and smaller sizes if they are "young." Higher designability thus reflects *the potential* for higher family size but does not necessarily imply it.

Another interesting observation is that older domains seem more designable. One may speculate that early protein evolution could have imposed more stringent constraints on domain designability either due to more challenging conditions (e.g. higher temperature) [39] or due to insufficient time to search effectively sequence space to make it possible to select viable sequences for less designable structures.

The findings presented here may have broad implications for our understanding of structural genomics as well as structure-function relationships and co-evolution. However, more quantitative evolutionary models are required to fully rationalize our findings. Further research along these lines may provide new insights into the genetic mechanisms underlying both neo-functionalization and the potential development of resistance to emerging diseases. These results provide an example of how fundamental physical principles can be statistically predictive in the biological Universe of protein folds and gene sequences.

**Methods**

*PDUG*

In order to build the PDUG, we use sequences from NRDB90[26] and all structural domains from HSSP[24]. We use BLAST[28] sequence homology to find all sequences in NRDB90 with more than 25% sequence identity to each HSSP domain. We combine that set of sequences into a single gene family. We then use cross-indexing between Swiss-Prot[25] and InterPro to find the set of all equilogs (sequences with the same function)[27] belonging to every gene family. We use those equilogs to reconstruct the FFS using Eq.2. (see Fig 1.) We use DALI[29] to make all pairwise structural comparisons and we build structural neighborhoods as described in the text and in Fig1. For this study we use Dali $Z_c=9$ as the cutoff value at which we consider two domains to be structural neighbors, although we believe that changing this value will not drastically alter the results, as evidenced by the correlation between domains and FFS (Fig 3a). We choose $Z_c=9$ because this level of structural divergence corresponds roughly to the superfold level of SCOP. Further justification of this threshold selection is given in Dokholyan et. al.[8].

An important issue in this paper is one of sequence weighting. The use of NRDB to exclude close sequence homologues ensures that we calculate sequence entropy by including far diverged sequences. The calculations of FFS provide another corroboration with the same result but a different weighting of sequences. Inclusion of all sequences from SWISS-PROT will introduce noise due to over-sequencing of some genes versus others and will not yield a sufficient approximation of entropy in sequence space.

*Designability:*

England and Shakhnovich showed recently [18] that for a large class of amino acid interaction potentials *B*, the free energy per monomer *f in sequence space* for a protein structure defined by its contact matrix (CM) *C* is given by

$$f = -\frac{1}{N}\sum_{n=2}^{\infty}(\text{Tr } C^n)a_n \quad (1)$$

where the weights $a_i$ are all positive functions which depend on the interaction energies *B*. The contact matrix C is defined as $C_{ij}=1$ if amino acids i and j are in contact and 0 otherwise. Definitions of contact may vary, but in this paper we use the standard cutoff of 7.5 angstroms between $C_\beta$ atoms ($C_\alpha$ for Gly). Elementary matrix algebra suggests that trace of high powers of a matrix is determined by its maximal eigenvalue. Thus, protein structures that have greater maximal eigenvalues of their contact matrices are expected to be more designable.

*Calculation of Variability in CD of Intra-family members*:

To calculate the variability of designability on Fig. 3 (error bars on the X axis) we considered all solved structures where the sequences are homologous to a representative domain on PDUG. We calculated the CD for all domains inside each sequence family where the number of homologous domains with resolved structures was larger than 2. For this calculation we used the domain boundaries that were delineated for the whole PDB[40] by Dietmann and Holm[24]. This resulted in consideration of over 34000 domains in approximately 3400 non-redundant representative homologous gene families. For each homologous sequence family, we calculated the standard deviation of CD for the structures belonging to that family. We then averaged all calculated standard deviations for gene families falling inside the gene family bin on Fig 3 and represented that quantity as the error bars on the CD axis.

*FFS*

In order to calculate functional entropy, we begin by combining all sequences into a set. We then match these sequences to InterPro[27] equilogs. We reconstruct the whole GO tree from the annotations of equilogs and calculate the number of equilogs of the family that are assigned a particular functional annotation, normalized by total number of annotations at each level (see Fig 1). We may thus calculate the average amount of information per annotation level needed to fully describe the function of each gene family using the following equation:

$$FFS = -\frac{1}{Max(L)} \sum_{l} \sum_{i \in \{\text{nodes on Level } l\}} p_i Log(p_i). \qquad (2)$$

Here, *Max(L)* is the maximal number of levels of annotation, the summation is taken over all levels $l$ and over all nodes $i$ filled by the gene family on the GO tree, and $p_i$ is the percentage of the family that is annotated with function $i$ (see Fig.1).

**Acknowledgements**

We are grateful to Jeremy England and Hooman Hennessey for their help as well as to Nikolay Dokholyan, Andrew Murray and Nick Grishin for fruitful discussions and critical readings of the manuscript and to NIH for support.

**References**

1. Finkelstein, A. V. & Ptitsyn, O. B. (1987). Why do globular proteins fit the limited set of folding patterns? *Prog Biophys Mol Biol* 50, 171-90.
2. Orengo, C. A., Todd, A. E. & Thornton, J. M. (1999). From protein structure to function. *Curr Opin Struct Biol* 9, 374-82.
3. Finkelstein, A. V., Gutin, A. M. & Badretdinov, A. (1995). Boltzmann-like statistics of protein architectures. Origins and consequences. *Subcell Biochem* 24, 1-26.
4. Koonin, E. V., Wolf, Y. I. & Karev, G. P. (2002). The structure of the protein universe and genome evolution. *Nature* 420, 218-23.
5. Teichmann, S. A., Chothia, C. & Gerstein, M. (1999). Advances in structural genomics. *Curr Opin Struct Biol* 9, 390-9.
6. Vitkup, D., Melamud, E., Moult, J. & Sander, C. (2001). Completeness in structural genomics. *Nat Struct Biol* 8, 559-66.
7. Yanai, I., Camacho, C. J. & DeLisi, C. (2000). Predictions of gene family distributions in microbial genomes: evolution by gene duplication and modification. *Phys Rev Lett* 85, 2641-4.
8. Dokholyan, N. V., Shakhnovich, B. & Shakhnovich, E. I. (2002). Expanding protein universe and its origin from the biological Big Bang. *Proc Natl Acad Sci U S A* 99, 14132-6.
9. Deeds, E. J., Dokholyan, N. V. & Shakhnovich, E. I. (2003). Protein Evolution within a Structural Space. *Biophys J*.
10. Qian, J., Luscombe, N. M. & Gerstein, M. (2001). Protein family and fold occurrence in genomes: power-law behaviour and evolutionary model. *J Mol Biol* 313, 673-81.
11. Huynen, M. A. & van Nimwegen, E. (1998). The frequency distribution of gene family sizes in complete genomes. *Mol Biol Evol* 15, 583-9.
12. Taverna, D. M. & Goldstein, R. A. (2000). The distribution of structures in evolving protein populations. *Biopolymers* 53, 1-8.
13. Li, H., Helling, R., Tang, C. & Wingreen, N. (1996). Emergence of preferred structures in a simple model of protein folding. *Science* 273, 666-9.
14. Miller, J., Zeng, C., Wingreen, N. S. & Tang, C. (2002). Emergence of highly designable protein-backbone conformations in an off-lattice model. *Proteins* 47, 506-12.
15. Govindarajan, S. & Goldstein, R. A. (1996). Why are some proteins structures so common? *Proc Natl Acad Sci U S A* 93, 3341-5.
16. Koehl, P. & Levitt, M. (2002). Protein topology and stability define the space of allowed sequences. *Proc Natl Acad Sci U S A* 99, 1280-1285.
17. Wolynes, P. G. (1996). Symmetry and the energy landscapes of biomolecules. *Proc Natl Acad Sci U S A* 93, 14249-55.
18. England, J. L. & Shakhnovich, E. I. (2003). Structural determinant of protein designability. *Physical Review Letters* in press.
19. Shakhnovich, E. I. (1998). Protein design: a perspective from simple tractable models. *Fold Des* 3, R45-58.

20. Lespinet, O., Wolf, Y. I., Koonin, E. V. & Aravind, L. (2002). The role of lineage-specific gene family expansion in the evolution of eukaryotes. *Genome Res* 12, 1048-59.
21. Manning, G., Plowman, G. D., Hunter, T. & Sudarsanam, S. (2002). Evolution of protein kinase signaling from yeast to man. *Trends Biochem Sci* 27, 514-20.
22. Bashford, D., Chothia, C. & Lesk, A. M. (1987). Determinants of a protein fold. Unique features of the globin amino acid sequences. *J Mol Biol* 196, 199-216.
23. Orengo, C. A., Pearl, F. M. & Thornton, J. M. (2003). The CATH domain structure database. *Methods Biochem Anal* 44, 249-71.
24. Dietmann, S., Park, J., Notredame, C., Heger, A., Lappe, M. & Holm, L. (2001). A fully automatic evolutionary classification of protein folds: Dali Domain Dictionary version 3. *Nucleic Acids Res* 29, 55-7.
25. Boeckmann, B., Bairoch, A., Apweiler, R., Blatter, M. C., Estreicher, A., Gasteiger, E., Martin, M. J., Michoud, K., O'Donovan, C., Phan, I., Pilbout, S. & Schneider, M. (2003). The SWISS-PROT protein knowledgebase and its supplement TrEMBL in 2003. *Nucleic Acids Res* 31, 365-70.
26. Holm, L. & Sander, C. (1998). Removing near-neighbour redundancy from large protein sequence collections. *Bioinformatics* 14, 423-9.
27. Apweiler, R., Attwood, T. K., Bairoch, A., Bateman, A., Birney, E., Biswas, M., Bucher, P., Cerutti, L., Corpet, F., Croning, M. D., Durbin, R., Falquet, L., Fleischmann, W., Gouzy, J., Hermjakob, H., Hulo, N., Jonassen, I., Kahn, D., Kanapin, A., Karavidopoulou, Y., Lopez, R., Marx, B., Mulder, N. J., Oinn, T. M., Pagni, M., Servant, F., Sigrist, C. J. & Zdobnov, E. M. (2000). InterPro--an integrated documentation resource for protein families, domains and functional sites. *Bioinformatics* 16, 1145-50.
28. Altschul, S. F., Madden, T. L., Schaffer, A. A., Zhang, J., Zhang, Z., Miller, W. & Lipman, D. J. (1997). Gapped BLAST and PSI-BLAST: a new generation of protein database search programs. *Nucleic Acids Res* 25, 3389-402.
29. Holm, L. & Sander, C. (1993). Protein structure comparison by alignment of distance matrices. *J Mol Biol* 233, 123-38.
30. Sonnhammer, E. L., Eddy, S. R., Birney, E., Bateman, A. & Durbin, R. (1998). Pfam: multiple sequence alignments and HMM-profiles of protein domains. *Nucleic Acids Res* 26, 320-2.
31. Baker, D. & Sali, A. (2001). Protein structure prediction and structural genomics. *Science* 294, 93-6.
32. Hegyi, H. & Gerstein, M. (2001). Annotation transfer for genomics: measuring functional divergence in multi-domain proteins. *Genome Res* 11, 1632-40.
33. Landau, L. D., Lifshi*t*s, E. M. & Pitaevski*i, L. P. (1978). *Statistical physics*. 3rd rev. and enl. / edit. Course of theoretical physics ; v. 5, 9 (Landau, L. D., Ed.), Pergamon Press, Oxford ; New York.
34. Grzybowski, B. A., Ishchenko, A. V., Shimada, J. & Shakhnovich, E. I. (2002). From knowledge-based potentials to combinatorial lead design in silico. *Acc Chem Res* 35, 261-9.
35. Ashburner, M., Ball, C. A., Blake, J. A., Botstein, D., Butler, H., Cherry, J. M., Davis, A. P., Dolinski, K., Dwight, S. S., Eppig, J. T., Harris, M. A., Hill, D. P., Issel-Tarver, L., Kasarskis, A., Lewis, S., Matese, J. C., Richardson, J. E.,

Ringwald, M., Rubin, G. M. & Sherlock, G. (2000). Gene ontology: tool for the unification of biology. The Gene Ontology Consortium. *Nat Genet* 25, 25-9.
36. Mirkin, B. G., Fenner, T. I., Galperin, M. Y. & Koonin, E. V. (2003). Algorithms for computing parsimonious evolutionary scenarios for genome evolution, the last universal common ancestor and dominance of horizontal gene transfer in the evolution of prokaryotes. *BMC Evol Biol* 3, 2.
37. Ponting, C. P. & Russell, R. R. (2002). The natural history of protein domains. *Annu Rev Biophys Biomol Struct* 31, 45-71.
38. Tiana, G., Shakhnovich, B. E., Dokholyan, N. & Shakhnovich, E. I. (2003). Imprint of Evolution on Protein Structures. *Submitted*.
39. England, J. L., Shakhnovich, B. E. & Shakhnovich, E. I. (2003). Natural selection of more designable folds: a mechanism for thermophilic adaptation. *Proc Natl Acad Sci U S A* 100, 8727-31.
40. Westbrook, J., Feng, Z., Chen, L., Yang, H. & Berman, H. M. (2003). The Protein Data Bank and structural genomics. *Nucleic Acids Res* 31, 489-91.

**Figure Legends**

**Fig. 1** A schematic picture of the scaled organization and intrinsic properties of the protein domain universe graph. The PDUG is built hierarchically, so that each level of evolutionary divergence can be considered independently. The domain structures are compared to each other (see Methods) and from this information the structural graph is created[8]. All the sequences from NRDB with more than twenty five percent identity to the original sequence of each domain on PDUG are collected into a gene family. All the equilogs (sequences with the same function)[27] matching the gene family are collected and used to create a probabilistic GO tree from which the FFS is calculated using Eq.2. As an example of how to build a structural neighborhood, consider the domain inside the blue rectangle, then all the domains with red rectangles are its structural neighbors.

**Fig. 2** An illustration of physical reasons for differences in designability between two structures. The balls schematically represent amino acids. Suppose that the interaction between the "red" amino acid and the "blue" amino acid is favorable and gives E = -1. The configuration on the left yields lower energy –4, compared with right structures where contribution from interactions between these amino acids is only –3. Thus the 4-loop in the left structure contributes more to the stability of the structure overall allowing more freedom to select the remaining part of the sequence to obtain overall stabilization of the structure, Similar considerations apply to 3-loops, 5-loops etc.

**Fig. 3**(a) The plot of the logarithm of average gene family size versus the structural contact density parameter calculated for the structures encoded by these sequences (as explained in Methods). Each point represents a bin in log (gene family size), with a step size of approximately 0.35. Each bin contains 100-250 families. Binning in (log) of gene

family sizes provides the advantage of having approximately equal number of gene families in each bin. The statistical correlation of the linear fit is R=0.95 with p<.001. The error bars on the CD axis represent the average deviation of CD inside each gene family averaged for all families belonging to the bin. (See methods) The error bars on the vertical axis correspond to the deviation of the mean number of members for each gene family inside the bin. (b). The correlation between the average CD of the structural neighborhood as defined on the PDUG (Fig1) and the log of the family sizes of all the sequences inside that neighborhood. Here, R=0.95 with p<.001. The error bars are calculated as described for (a).

**Fig 4**. The correlation between CD and functional flexibility score (FFS) of the gene family calculated via equilogs using Eq.1 This is evidence that structural determinant of designability, CD serves as a direct influence on the number of functions that a gene family does, with linear fit correlation R=0.97. Each datapoint represents a bin in FFS, with step 0.1 containing 50-200 families. The datapoints represent the average CD over all gene families represented in a FFS bin.

**Fig.5** Scatter plot of gene family size vs CD for all PDUG domains. Red data points correspond to 108 ''most ancient'' domains defined as all domains shared by all prokaryotic proteomes. (see supplementary information for the list of 108 LUCA domains).Note that most ancient domains are statistically more designable (higher CD, difference of means .48, P-value less than 1e-14) and that at the same CD, their families are more populated, on average 116 more members in each family. (P-value less than 1e-14). The range of sequence family size and CD for LUCA domains and the very large

difference between LUCA family size and non-LUCA indicates that the relative addition to this data from a sampling bias due to inclusion in all 59 genomes is small.

**Figures**

Fig 1.

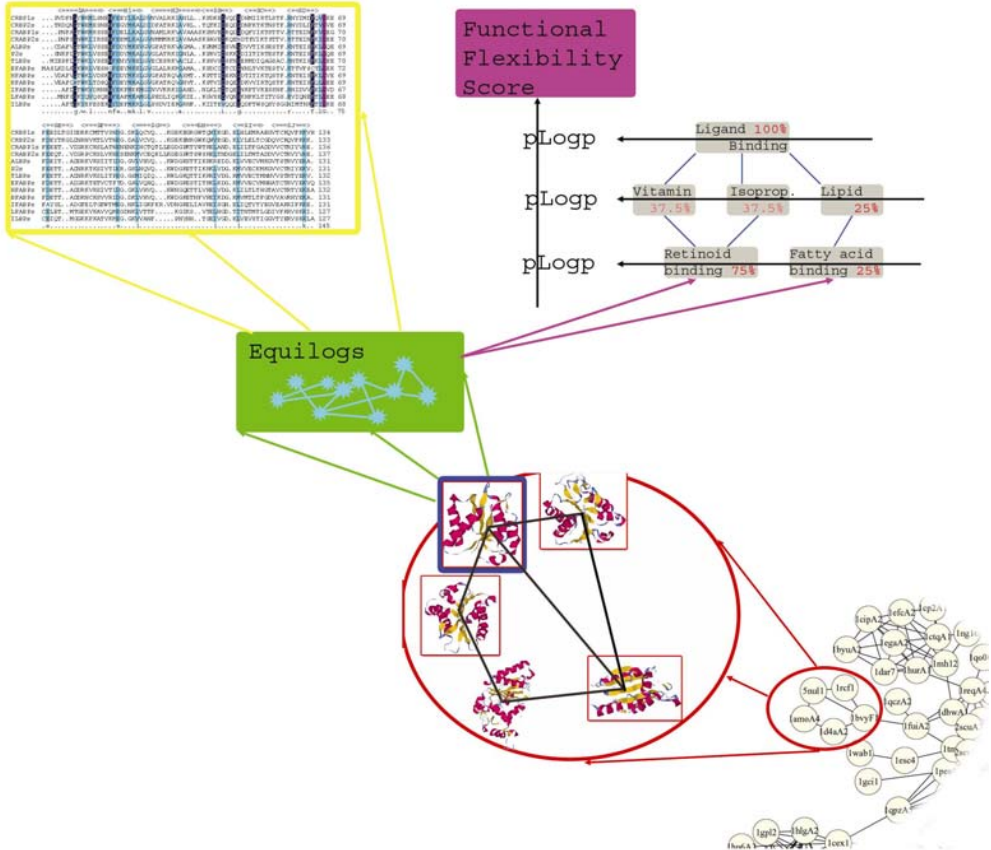

Fig 2.

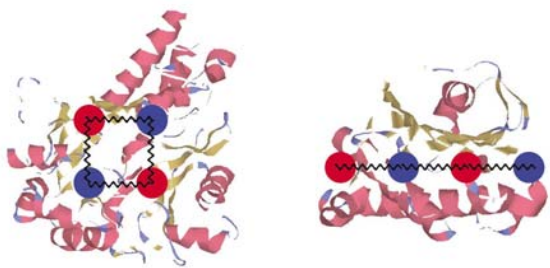

Fig 3a.

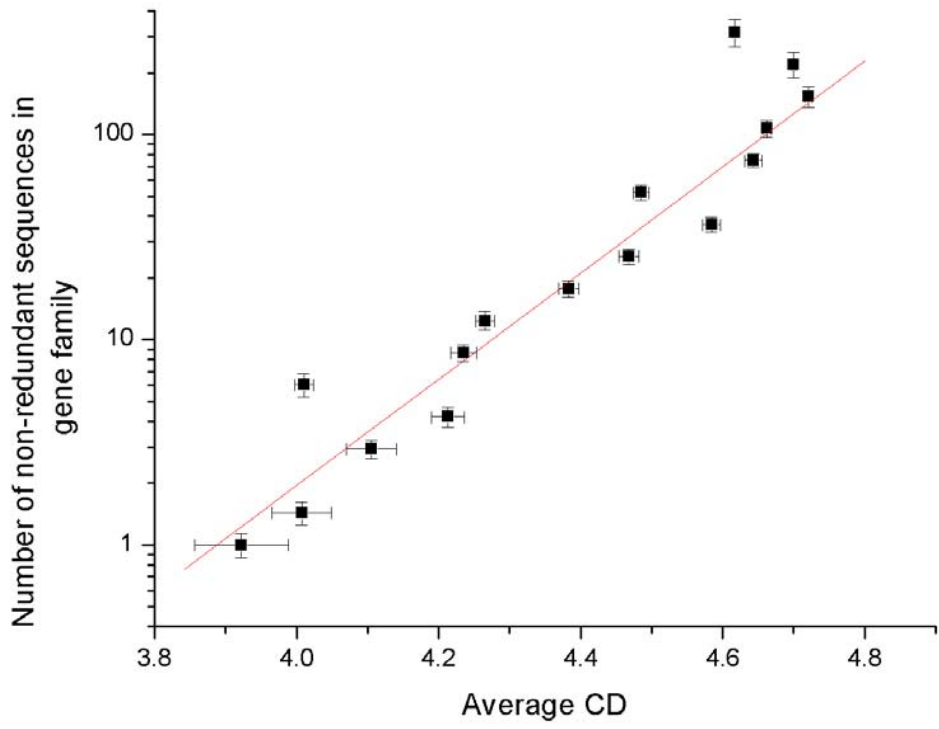

Fig 3b.

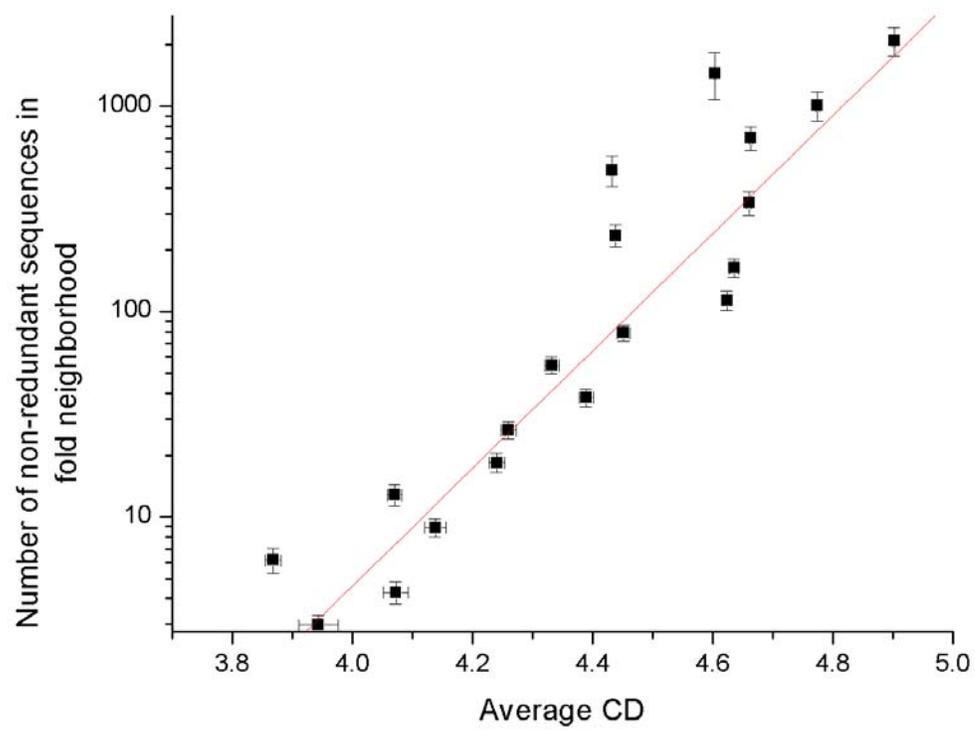

Fig.4

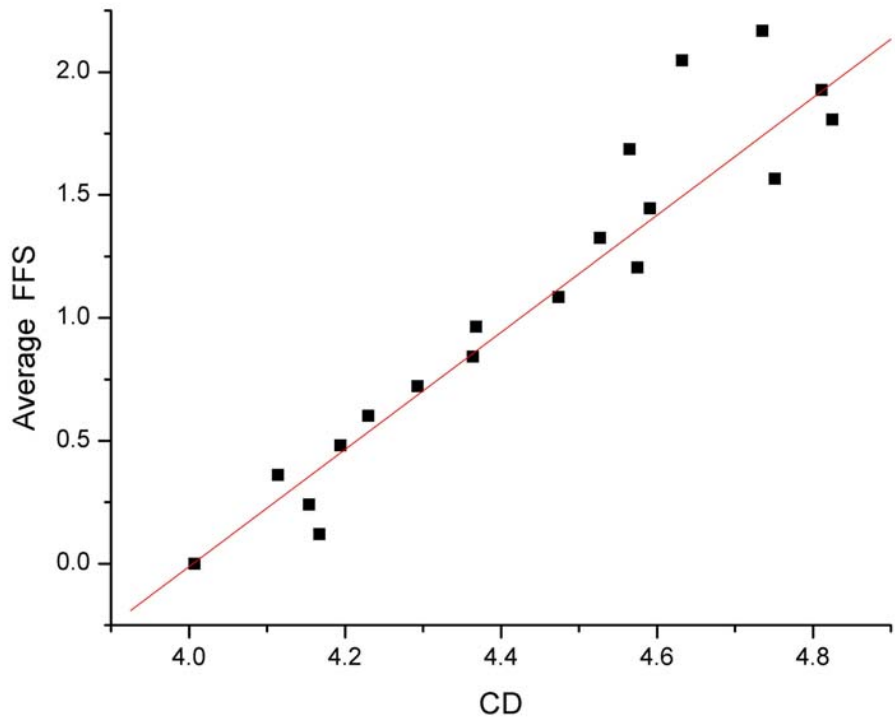

Fig.5

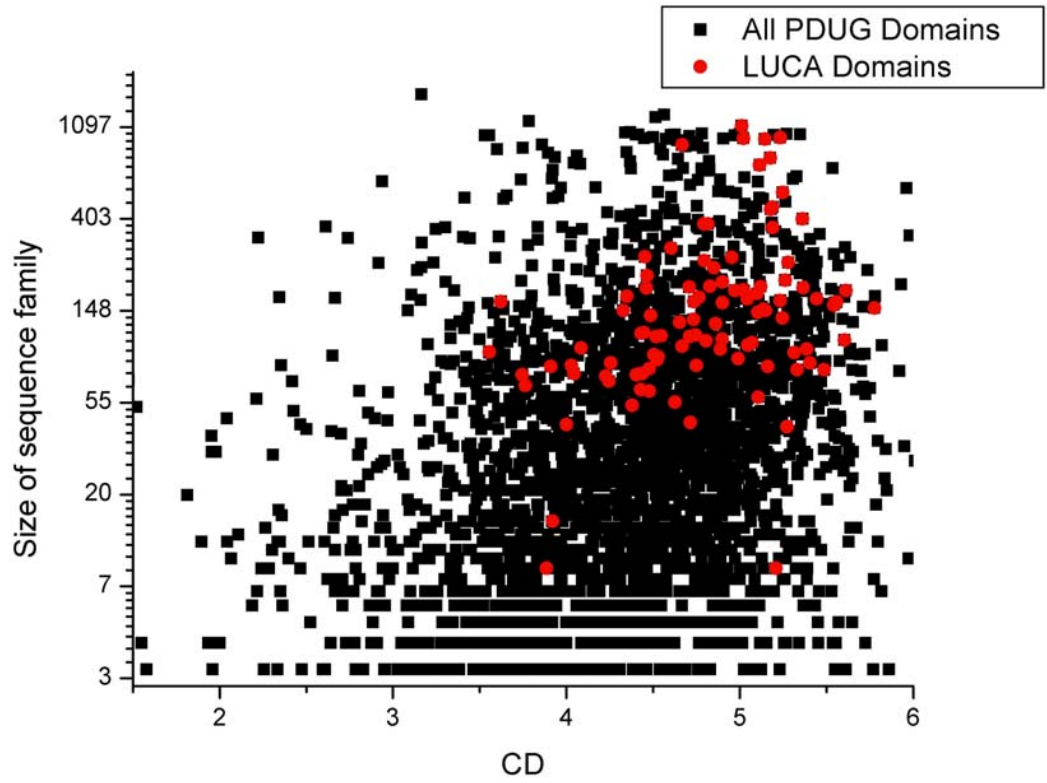